\documentclass[11pt,prd,preprint,superscriptaddress,showkeys]{revtex4}
\usepackage{amsmath}
\usepackage[dvips]{graphicx}
\newcommand{\be}{\begin{equation}}
\newcommand{\ee}{\end{equation}}
\newcommand{\ben}{\begin{eqnarray}}
\newcommand{\een}{\end{eqnarray}}

\begin{document}

\title{On the consistency of warm inflation in the presence of viscosity}
\date{\today}
\author{S. del Campo\footnote{E-mail: sdelcamp@ucv.cl}}
\affiliation{Instituto de F\'{\i}sica, Pontificia Universidad
Cat\'{o}lica de Valpara\'{\i}so, Av. Universidad 330, Campus
Curauma, Valpara\'{\i}so , Chile}

\author{R. Herrera\footnote{E-mail: ramon.herrera@ucv.cl}}
\affiliation{Instituto de F\'{\i}sica, Pontificia Universidad
Cat\'{o}lica de Valpara\'{\i}so, Av. Universidad 330, Campus
Curauma, Valpara\'{\i}so , Chile}

\author{D. Pav\'{o}n\footnote{E-mail: diego.pavon@uab.es}}
\affiliation{Departamento de F\'{\i}sica, Facultad de Ciencias,
Universidad Aut\'{o}noma de Barcelona, 08193 Bellaterra
(Barcelona), Spain}

\author{J.R.Villanueva\footnote{E-mail: jose.villanueva.l@mail.ucv.cl}}
\affiliation{Instituto de F\'{\i}sica, Pontificia Universidad
Cat\'{o}lica de Valpara\'{\i}so, Av. Universidad 330, Campus
Curauma, Valpara\'{\i}so , Chile}

\begin{abstract}
This paper studies the stability of warm inflationary solutions
when the viscous pressure is taken into account. The latter is a
very natural and physically motivated ingredient of warm inflation
and is seen to widen the stability range of warm inflation. The
spectral index parameters, $n_{s}$, $n_{T}$, and their ratio  are
derived. The corresponding WMAP7 data are used to fix some
parameters of the model. Two specific examples are discussed in
detail: (i) a potential given by $V(\phi, T) = v_{1}(\phi)\,  + \,
v_{2}(T)$, and (ii) a potential of the form $V(\phi, T) = \alpha
\, v_{1}(\phi) \, v_{2}(T)$. In both cases, the viscosity has
little impact on the said ratio.

\end{abstract}


\maketitle

\section{Introduction}
Nowadays it is widely admitted that our Universe experienced a
very early stage of accelerated expansion (inflation) which
served, among other things, to produce the seeds that, in the
course of the subsequent eras of radiation and matter dominance,
developed into the cosmic structures (galaxies and clusters
thereof) that we observe today \cite{lyth1,lyth2}. Broadly
speaking, we might say there are two main competing scenarios of
slow roll inflation: cold inflation \cite{linde1,linde2}, that
ends in a very short and violent phase of reheating, and warm
inflation \cite{berera-fang,berera}, that does not necessitate the
latter phase as the inflaton field decays into radiation at a
sufficient (damping) rate, say $\Gamma$, during the accelerated
expansion. This essential feature enables the Universe to smoothly
proceed into the radiation phase, indispensable for the big bang
nucleosynthesis. A further characteristic of warm inflation is
that the radiation temperature satisfies $T > H$. As a
consequence, thermal fluctuations result bigger than quantum
fluctuations whereby the former likely are the main origin of
cosmic structures \cite{taylor2000}.

Any inflationary model -whether ``cold" or ``warm"- must fulfill
the requirement of stability; that is to say, its inflationary
solutions ought to be attractors in the solution space of the
relevant cosmological solutions. It means, in practice, that the
scalar field, $\phi$, must approach an asymptotic attractor
characterized by $\dot{\phi} \simeq - (\partial V/\partial
\phi)\,(3H)^{-1}$ in cold inflation, and $\dot{\phi} \simeq -
 (\partial V/\partial \phi)(3H + \Gamma)^{-1}$ in warm inflation (see e.g.
\cite{salopek,parsons}). This ensures that the system will stay
sufficiently near to the slow-roll solution for many Hubble times.
Here $V$ denotes the scalar field potential and $H$ the Hubble
expansion rate.

In the case of warm inflation the conditions for stability have
been considered by de Oliveira and Ramos \cite{henrique} and,
recently, more fully by Moss and Xiong \cite{moss2008} who allowed
the scalar potential and the damping rate to depend not only on
the inflaton field but on the temperature of the radiation gas as
well. This automatically introduces two further slow-roll
parameters and renders the conditions for a successful warm
inflationary scenario even less restrictive.

The aim of  this paper is to take a further step in the analysis
of the stability of warm inflationary solutions by considering the
presence of massive particles and fields in the radiation fluid as
well as the existence of a the viscous pressure, $\Pi$, associated
to the resulting mixture of heavy and light particles. This novel
quantity is well motivated on physical grounds. It can represent
either a genuine viscous pressure (as in radiative fluids
\cite{weinberg1,weinberg2}), or an effective negative pressure
linked to the production of particles  by the inflaton field, or
the decay of the heavy fields into light fields (see, e.g.
\cite{zeldovich,bei-lok,barrow,pla}), or both. Whatever the case,
it may significantly influence the dynamics of warm inflation. In
particular, the inflationary region occupies a wider section of
the phase space associated to the autonomous system of
differential equations when $\Pi \neq 0$ than otherwise
\cite{josepedro}, and its effect on the matter power spectrum (of
a few per cent) will be possibly detected in future experiments
\cite{srd}. Moreover, the adiabatic index $\gamma$ of the mixture
will not necessary be $4/3$ but some value in the range $1 <
\gamma < 2$, that will depend on the proportion between heavy and
light particles in the mixture.

This paper is organized as follows. Section  II presents the basic
equations for warm inflation with viscosity. Section III analyzes
the stability for inflationary solutions. Section IV determines
the power spectrum produced by the thermal fluctuations.
Corrections to the spectral index are considered when the
viscosity comes into play. Section V applies the full theory to
two specific examples differing in the type of potential but with
identical forms for the viscous pressure and damping rate,
$\Gamma$. The seven year WMAP data are used to constrain the
model. Finally, Section VI summarizes and discusses our findings.
Hence forward a coma means partial derivative. We use units in
which $c = \hbar = 8 \pi G = 1$.

\section{Basic equations}
We consider a spatially flat, homogeneous and isotropic universe
dominated by an scalar field $\phi$ and radiation, of temperature
$T$, endowed with a viscous pressure $\Pi = -3\, \zeta H$, where
$\zeta$ is the semi-positive definite coefficient of bulk
viscosity. A potential $V(\phi, T)$ and the damping rate
$\Gamma(\phi, T)$ dictate the evolution of the  field so that the
corresponding Klein-Gordon equation assumes the form
\be \ddot \phi + (3 H +\Gamma)\dot\phi + V_{,\phi}=0 \, .
\label{1}
\ee

In this scenario, the Friedmann equation reads
\be 3 H^{2}= \rho_{\phi} \, + \, \rho_{\gamma}\, ,
\label{2}
\ee
where $\rho_{\phi} = (1/2) \dot{\phi}^{2} \, + \, V(\phi, T)$
\noindent and $\rho_{\gamma} = Ts$, denote the energy densities of
the scalar field (the inflaton) and radiation, respectively. Here
\be s
\simeq - V_{,T} \, , \label{4}
\ee
\noindent is the entropy density. This expression follows from the
thermodynamical relation $s = - \partial f/\partial T \,$, when
the Helmholtz free-energy, $f = (1/2)\dot{\phi}^{2} \, + \,
V(\phi, T) \, + \, \rho_{\gamma} \, - \, Ts$, is dominated by the
potential.

Since for an expanding Universe the viscous pressure is
necessarily negative the total pressure
\be
p = \frac{1}{2}\dot \phi ^{2} \, - \, V(\phi, T) \, + \,
(\gamma - 1)T s \, + \, \Pi,
\label{5}
\ee
\noindent is lower than in the absence of viscosity. Obviously,
the immediate effect of $\Pi$ is to further accelerate the
expansion. Clearly, the adiabatic coefficient, $\gamma$, must
slowly vary with expansion but, for simplicity sake, we will
consider it constant.

The conservation of the stress-energy can be expressed as
\be
T \dot s \, + \, 3 H(\gamma T s \, + \, \Pi)= \Gamma \dot \phi
^{2} \, .
\label{6}
\ee
Making $u = \dot \phi$ and introducing the dimensionless strength
dissipation parameter, $Q = \Gamma/(3H)$, the slow roll equations
take the form
\be
u = \frac{- V_{,\phi}}{3 H (1 + Q)}, \qquad T s = \frac{Q
u^{2} \, + \, 3H \zeta}{\gamma}, \qquad 3 H^{2} = V(\phi, T)\, .
\label{7}
\ee

The number of e-folds is given by
\be N = \int_{\phi_{*}}^{\phi_{end}}\frac{3 H^{2} (1 \, + \,
Q)}{V_{,\phi}(\phi, T)} \, d\phi \,, \label{9.1} \ee
where $\phi_{*}$ and $\phi_{end}$ denote the value of the inflaton
field when the perturbations cross the horizon (i.e., when $k = a
H$) and at the end of inflation, respectively.

Finally, in warm inflation the customary slow roll parameters,
\be
\epsilon =\frac{1}{2}\left( \frac{V_{,\phi}}{V} \right)^{2}\,
, \quad \eta = \frac{V_{,\phi \phi}}{V}\, , \qquad \beta =
\frac{V_{,\phi} \Gamma_{,\phi}}{V \Gamma}\, ,
\label{10}
\ee
\noindent are supplemented by two others,
\be
b = \frac{T V_{,\phi T}}{V_{,\phi}}\, , \qquad c = \frac{T
\Gamma_{,T}}{\Gamma}\, ,
\label{101}
\ee
that gauge the temperature dependence of the potential and the
damping rate, respectively.

\section{Stability analysis}
We are interested in finding the impact of the viscous pressure,
$\Pi$, on the stability of the slow roll. This immediately implies
that we must focus on the strong regime (i.e., $Q \gg 1$) as in
the weak regime ($Q \ll 1$) viscosity will be practically
non-existent \cite{srd}.

To find the conditions for the validity of the slow roll
approximation, we perform a linear stability analysis to see
whether the system remains close to the slow roll solution for
many Hubble times. In cold inflationary scenario, the slow roll
equation is of first order in the time derivative. Choosing the
inflaton field as independent variable, the conservation equations
(\ref{1}) and (\ref{6}) can be written as first order equations in
the derivative with respect to $\phi$, indicated by a prime,
\be
x' = F(x)\, ,
\label{11}
\ee
\noindent where
\be
x = \left(
          \begin{array}{c}
            u \\
            s \\
          \end{array}
        \right).
        \label{12}
        \ee

Thus, the system (\ref{1}), (\ref{6}) becomes
\be
u'= -3H\, - \, \Gamma -\, V_{,\phi}u^{-1}\, ,
\label{13}
\ee
\be
s' = -3H \gamma s u^{-1}\, - \, 3H \Pi(T u)^{-1}\, + \,
T^{-1}\Gamma u \, .
\label{14} \ee
\noindent Here the Hubble rate and temperature are determined by
(\ref{2}) and (\ref{4}), respectively.

Taking a background $\overline{x}$ which satisfies the slow roll
equations (\ref{7}), the linearized perturbations satisfy

\be
\delta x'= M(\overline{x}) \delta x-\overline{x}'\,
,
\label{15}
\ee
\noindent where
\be
M=\left(
        \begin{array}{cc}
          A & B \\
          C & D \\
        \end{array}
      \right),
      \label{16}
      \ee
is the matrix of first derivatives of $F$ evaluated at the slow
roll solution. Linear stability demands that its determinant be
positive and its trace negative.

The matrix elements read,
\be
A = \frac{H}{u}\left\{-3(1+Q)\, - \,
\frac{\epsilon}{(1+Q)^{2}}\right\}\, ,
\label{17}
\ee
\be
B = \frac{H}{s}\left\{-c \, Q-\frac{Q}{(1+Q)^{2}}\epsilon \, +
\, b (1+Q)  \right\}\, ,
\label{18}
\ee
\be
C = \gamma\frac{H s}{u^{2}}\left(6-\frac{\epsilon}{(1+Q)^{2}}
\right) \left\{1+\frac{\Pi}{\gamma^{2} \rho_{\gamma}}\left(\frac{6
(1+Q)^{2}-2\epsilon}{6 (1+Q)^{2}-\epsilon} \right) \right\}\, ,
\label{19}
\ee

\be D = \gamma\frac{H}{u}\left(c- 4 -\frac{Q \epsilon}{\gamma^{2}
(1+Q)^{2}} \right)\, + \, \frac{H \Pi}{u \gamma \rho_{\gamma}}
\left\{c-\frac{Q \epsilon}{\gamma^{2} (1\, + \, Q)^{2}}\, + \,
\frac{3 \Pi}{2 \gamma^{2} \rho_{\gamma}} \right\}\, .
\label{20}
\ee


\noindent In the strong regime ($Q \gg 1$), the determinant and
trace of $M$ assume the comparatively simple expressions
\be
\det M = \frac{3 \gamma Q H^{2}}{u^{2}} \left(4 - 2 b + c+
(c-2 b )  \frac{\Pi}{\gamma^{2} \rho_{\gamma}} \, - \, \frac{3}{2}
\frac{\Pi^{2}}{\gamma^{4} \rho_{\gamma}^{2}} \right)\, ,
\label{23}
\ee
\noindent and
\be \textrm{tr} M = \frac{H}{u}\left\{-3 Q \, + \, \gamma(c-4) \,
+ \, \frac{\Pi} {2 \gamma^{3} \rho_{\gamma}}\left(2 \gamma^{2} c
\, + \, 3 \frac{\Pi}{\rho_{\gamma}} \right)\right\}\, .
 \label{24}
 \ee

Sufficient conditions for stability are that $M$ varies slowly and
that
\be |c| \leq \frac{4 - 3\sigma^{2} /2}{1 +  \sigma} - 2 b\, ,
\qquad b \geq 0 \, , \label{27} \ee
\noindent where $\sigma \equiv \frac{\Pi}{\gamma^{2}
\rho_{\gamma}}$. Upon these conditions  the determinant results
positive and the trace negative, implying stability of the
corresponding solution. Expression (\ref{27}.1) generalizes Eq.
(27) of Moss and Xiong \cite{moss2008}.

Since the  chosen background is not an exact solution of the
complete set of equations, the forcing term in equation (\ref{15})
depends on $\overline{x}'$, and will be valid only if
$\overline{x}'$ is small. The size of $\overline{x}'$ depends on
the quantities $\dot u/(H u)$ and $\dot s/(H s)$. From the time
derivative of (\ref{7}.3) we obtain
\be
\frac{\dot H}{H^{2}}=-\frac{\epsilon}{1 + Q}\, .
\label{28}
\ee
\noindent Combining this with the other slow-roll equations,
(\ref{7}.1) and (\ref{7}.2), we get
\be \frac{\dot u}{H u}= \frac{1}{\Delta}\left[- \frac{c[A  (1 \, +
\,  Q)\, - \, B  Q] \, - \, 4}{1 \, + \, Q} \epsilon \, + \,
\frac{4 Q}{1 \, + \, Q}\beta \, +\,  (A c \, - \, 4)\eta - \frac{3
(1 \, + \, Q) c}{1 \, - \,  f} b \right]\, , \label{29} \ee
\noindent and
\be \frac{\dot s}{H s}= \frac{3}{\Delta}\left[\frac{A  (3 \, + \,
Q)\, - \, B (1 \, + \, Q)}{1 \, + \, Q} \epsilon \, + \, \frac{Q
\, - \, 1}{1 \, + \, Q} A \beta \, -\, 2 A \eta - \frac{(1 + Q)[A
c (Q - 1) + Q + 1] c}{(1 - f) Q} b \right]\, , \label{30} \ee
\noindent where \be \Delta = 4 (1 + Q) + A c (Q - 1)\, , \qquad
 A = \frac{\rho_{\gamma} + \gamma^{-1} \Pi}{\rho_{\gamma} \, -
\, \kappa \Pi}\, , \qquad B = \frac{\Pi}{\rho_{\gamma} \, - \kappa
\Pi}\, , \label{33} \ee \be f = -\frac{3}{2}\frac{(1 +
Q)^{2}}{Q}\frac{\zeta}{\gamma H \epsilon}\, , \qquad \kappa =
\rho_{\gamma}\frac{\zeta_{,\rho_{\gamma}}}{\zeta} \,
.\label{34}\ee

Notice that when  $\Pi \rightarrow 0$, one has  that $A\rightarrow
1$, $B\rightarrow 0$, $f\rightarrow 0$, and therefore the
equations (\ref{28})-(\ref{30}) reduce to the corresponding
expressions in \cite{moss2008}. Obviously, the value of the
parameter $\kappa$ in this limit depends on the specific
expression of the viscosity coefficient, $\zeta$; but it does not
alter the value of $B$ in the said limit. In this limit, the
$\kappa$ parameter could take any value depending of the model.
Its value does not affect the $\Pi \rightarrow 0$ limit.

\section{Density fluctuations}
The thermal fluctuations produce a power spectrum of scalar
density fluctuations of the form \cite{moss2008}
\be
\mathcal{P}_{s} = \frac{\sqrt{\pi}}{2}\frac{H^{3}
T}{u^{2}}\sqrt{1 \, + \, Q}
\, . \label{36}
\ee
Note that the power spectrum of fluctuations in inflationary
models where the friction coefficient  depends also on the
temperature, i.e., $\Gamma=\Gamma(\phi,T)$, was considered
recently in Ref.\cite{Graham:2009bf}.

We calculate the spectral index by means of
\be
n_{s} - 1 = \frac{\mathcal{\dot P}_{s}}{H \mathcal{P}_{s}}\,
.
\label{37}
\ee
\noindent By virtue of the equations (\ref{28})-(\ref{30}), we
obtain
\be
n_{s}-1=\frac{p_{1}\epsilon+p_{2}\beta+p_{3}\eta+p_{4}b}{\Delta}\,
,\label{38}\ee
\noindent where the $p_{i}$ coefficients are given by
\be p_{1}=-\frac{10(2 \, + \, Q)-A(\, 3\, +5\, c\, +\, Q)\, +B\,
(1\, +\, Q\, +\, (5c/2) Q)}{1\, +\, Q}\, ,\label{39}\ee \be
p_{2}=\frac{A(Q\, -\, 1)\, -\, 10Q}{1\, +\, Q},\label{40}\ee \be
p_{3}=\frac{8(1\, +\, Q)-A(2\, +\, 2c\, +2Q\, +\, 3c Q)}{1\, +\,
Q},\label{41}\ee \be p_{4}=\frac{3(1\, +\, Q)[1\, +\, (1+5c/2
)Q]}{(1\, - \, f)Q}\, .\label{42}\ee

For $Q \gg 1$, and assuming $c$ of order unity, the $p_{i}$
coefficients reduce to

\be p_{1}= -10+A-B(1+5c/2)\, , \quad p_{2}=A-10, \quad
p_{3}=8-A(2+3c)\, , \quad p_{4}=\frac{3Q(1+5c/2)}{(1-f)} \, ,
\label{43} \ee
\noindent and $\Delta = Q(4+A c)$. Therefore (\ref{38}) becomes
\be
n_{s}-1=-\frac{10-A+B(1+5c/2)}{(4+A c)Q}\, \epsilon \, -\,
\frac{10-A}{(4+A c)Q}\, \beta \, + \,  \frac{8-A(2+3c)}{(4+A
c)Q}\, \eta \, + \,  \frac{3(1+5c/2)}{(4+A c)(1-f)}b\, .
\label{47}
\ee

The tensor modes happen to be the same as in the cold inflationary
models \cite{moss2008}, i.e.,
\be
\mathcal{P}_{T} = H^{2}\, ,
\label{53}
\ee

\noindent and the corresponding spectral index is
\be
 n_{T} \, - \, 1 = -\frac{2}{1+Q}\epsilon\, .
\label{54}
\ee

With the help of of (\ref{53}), (\ref{36}) and (\ref{7}.1) the
tensor-to-scalar amplitude ratio can be written as

\be r = \frac{2 \, V_{,\phi}(\phi, T)}{9 \sqrt{\pi} \, H^{3} \, T
(1 \, + \,  Q)^{5/2}}\, . \label{54.1} \ee
The  recent WMAP seven-year results imply the upper-bound $r<0.36$
(95$\%$ CL) \cite{Larson:2010gs} on the scalar-tensor ratio.
Below, shall make use of this bound to set constraints on the
parameters of our models.

\section{Examples}
In this section we apply the formalism of above to two specific
cases sharing the same damping and bulk viscosity coefficient but
differing in the potential. In general, the damping coefficient
may be written as

\be \Gamma(\phi, T)= \Gamma_{0}
\left(\frac{\phi}{\phi_0}\right)^{m}\;\left(\frac{T}{\tau_0}\right)^{n},
\label{56} \ee

\noindent with $n$ and $m$ real numbers and $\phi_0$, $\tau_0$,
and $\Gamma_{0}$ some  nonnegative constants. The damping term has
a generic form given approximately by $ \Gamma \sim g^4\phi^2
\tau$, where $g$ is the coupling constant \cite{HMB2004}. From
Ref.\cite{Hoso} the damping term, $\tau=\tau(\phi,T)$, is related
to the relaxation time of the radiation and for the models with an
intermediate particle decay, $\tau =\tau(\phi)$ is linked to the
lifetime of the intermediate particle. Different choices of $n$
and $m$ have been adopted. For instance the case $n = m = 0$ was
considered by Taylor and Berera \cite{taylor2000}, whereas the
choice $m = 2$, $n = -1$ corresponds to the damping term first
calculated by Hosoya \cite{Hoso}. This expression slightly differs
from those in \cite{HMB2004,zhang2009}, where a single index
rather than two was considered.

As for the bulk viscosity coefficient we use the general
expression
\be
\zeta = \zeta_{0}\, \rho_{\gamma}^{\lambda} \, ,
\label{zeta}
\ee
where $\zeta_{0}$ is a positive semi-definite constant and
$\lambda$ an integer that may take  any of the two values:
$\lambda=1/2$, i.e., $\zeta\propto\rho_{\gamma}^{1/2}$\cite{1}
 (see also Ref.\cite{2}) and $\lambda=1$, i.e.,
$\zeta\propto\rho_{\gamma}$ \cite{srd}.

\subsection{First example: a separable potential}
We begin by considering the comparatively easy separable potential
\cite{moss2008}
\be
V(\phi, T)= v_{1}(T) \,+ \, v_{2}(\phi)\, ,
\label{55}
\ee
\noindent where $v_{1}(T) = -\frac{\pi^{2}}{90}g_{*}T^{4} -
\frac{1}{12}m_{\phi}^{2} T^{2}$ and $v_{2}(\phi)=
\frac{1}{2}m_{\phi}^{2}\phi^{2}$.

In this case, the strength of dissipation parameter and the slow
rolls parameters are
\be Q = \frac{\Gamma(\phi,T)}{\sqrt{3 (v_{1} + v_{2})}}\, ,
\label{57} \ee
\noindent and
\be
\epsilon = \frac{m^{4}_{\phi}\phi^{2}}{2 V^{2}}\, ,\qquad
\beta = \frac{m \, m^{2}_{\phi}}{V}\, , \qquad \eta = \frac{
m^{2}_{\phi}}{V}\, , \qquad b=0\, , \qquad c = n\, \,
\label{59}
\ee
respectively.

\noindent Recalling that $\rho_{\gamma} = - T \, V_{,T}$, we
obtain the following expression for the radiation energy density
\be
\rho_{\gamma} = \frac{2 \pi^{2}}{45}g_{*}T^{4} \, + \,
\frac{1}{6}m_{\phi}^{2} T^{2}\, .
\label{60}
\ee
To get a lower limit to the friction term in the presence of
viscosity, we follow the method of \cite{moss2008}. We cast Eq.
(\ref{36}) in the form
\be
\mathcal{P}_{s} \approx \frac{T^{4}}{u^{2}}\frac{H^{3}
}{T^{3}}(1 + Q)^{1/2}\, ,
\label{61}
\ee
\noindent and use (\ref{7}.2) together with (\ref{56}) to arrive
to the cubic equation
\be
4 X^{3} - g_{2} X - g_{3} = 0\, ,
\label{62}
\ee
\noindent with
\be g_{2}=-\frac{12 \zeta \mathcal{P}_{s}}{Q^{*} T^{3}}\, , \qquad
g_{3}=\frac{8 \pi^{2} \gamma}{45}\frac{g_{*}
\mathcal{P}_{s}}{Q^{*}}\ , \label{63} \ee
where $Q^{*} \equiv Q \, \sqrt{1\, + \, Q}$ and  $X \equiv H /T$.

The only real solution of the Eq. (\ref{62}) is found to be
\be X = Z \,  \sinh \left(\theta \right),  \label{64} \ee
\noindent with
\be
Z = 2 \, \sqrt{\frac{\zeta \mathcal{P}_{s}}{Q^{*} T^{3}}}\, ,
\quad {\rm and}  \qquad \theta =  \frac{1}{3}
\sinh^{-1}\left(\frac{\pi^{2} \gamma}{45}\sqrt{\frac{g_{*}^{2}
Q^{*} T^{9}}{\zeta^{3} \mathcal{P}_{s}}}\right)\, .
\label{65}
\ee

Now, the condition for warm inflation, $T > H$ (i.e., $X < 1$),
translates into
\be
Q > g_{*} \mathcal{P}_{s} \left[1 - \frac{3}{2}
\left(\frac{45}{\pi^{2} \gamma g_{*}}\right)
\frac{\zeta}{T^{3}}\right]\, .
\label{67}
\ee
\noindent Note that, as $\zeta \geq 0$, the viscous pressure
augments the range of $Q$, and that the condition for warm
inflation without viscosity, $Q > g_{*} \, \mathcal{P}_{s} $, is
trivially recovered in the limit $\zeta \rightarrow 0$.

Using (\ref{zeta}) and (\ref{60}) the inequality (\ref{67}) can be
written as

\be
Q > g_{*} \mathcal{P}_{s} \left[1 - \frac{3 \zeta_{0}}{\gamma}
\left(\frac{2 \pi^{2} g_{*}}{45}\right)^{\lambda - 1} T^{4 \lambda
- 3}\right]\, .
\label{68}
\ee
\noindent Since $Q > 0$ an upper bound on $\zeta_{0}$ follows,
namely,
\be
\zeta_{0} < \zeta_{max}^{(1)} \equiv \frac{\gamma}{3}
\left(\frac{2 \pi^{2} g_{*}}{45}\right)^{1- \lambda} T^{3 - 4
\lambda}\, .
\label{69}
\ee
\begin{figure}[th]
\includegraphics[width=6.5in,angle=0,clip=true]{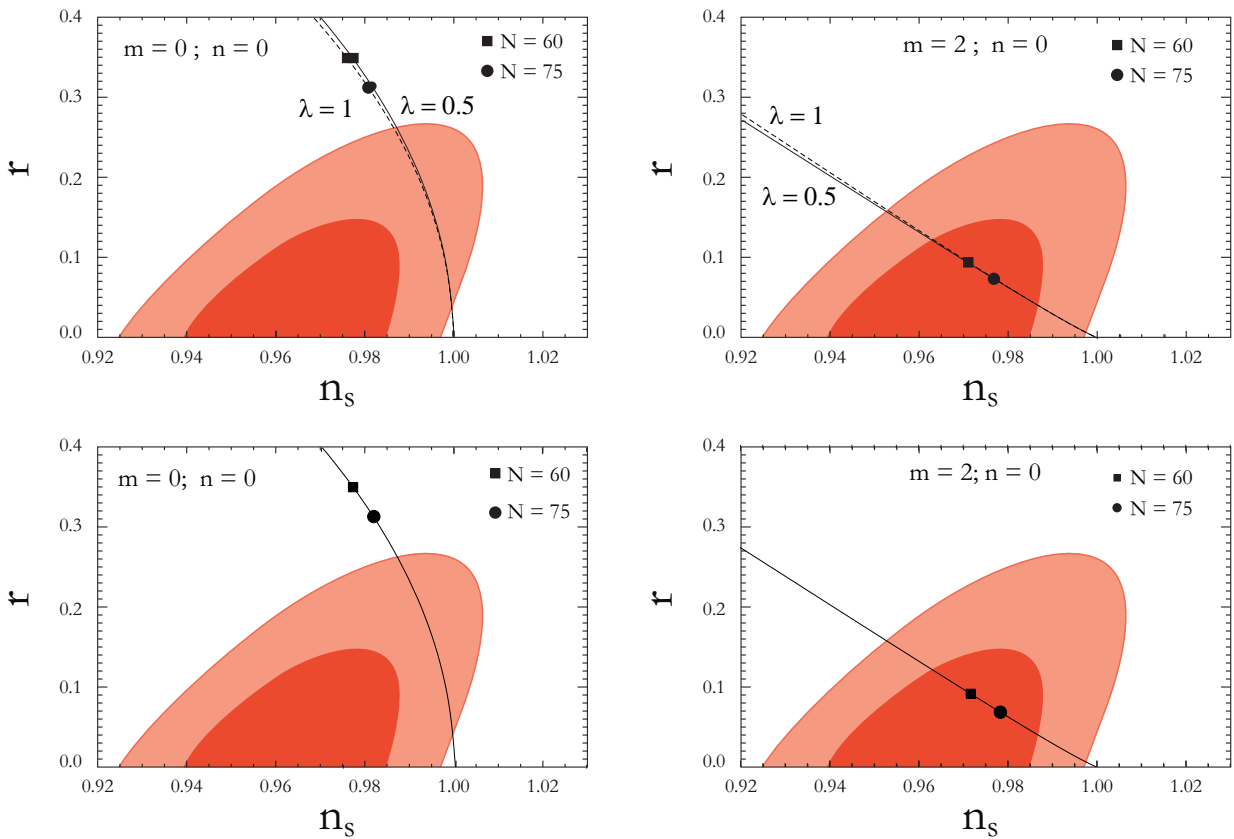}
\caption{Top row of panels: Plot of  the tensor-scalar ratio $r$
as a function of the spectral index $n_{s}$, for two values of the
$\lambda$ parameter in the case of example 1 (i.e., potential
(\ref{55})). Bottom row: Same as the top row but assuming no
viscosity ($\zeta_{0} = 0$). In each panel  the 68$\%$ and 95$\%$
confidence levels set by seven-year WMAP experiment are shown. The
latter places severe limits on the tensor-scalar ratio
\cite{Larson:2010gs}.} \label{fig1}
\end{figure}

Figure \ref{fig1} depicts the dependence of the tensor-scalar
ratio, $r$, on the spectral index, $n_s$, for the  model given by
Eqs. (\ref{56}), (\ref{zeta}), and (\ref{55}) when $\lambda=0.5$
and when $\lambda=1$. From Ref. \cite{Larson:2010gs},
two-dimensional marginalized constraints (68$\%$ and 95$\%$
confidence levels) on inflationary parameters $r$ and $n_s$, the
spectral index of fluctuations, defined at $k_0$ = 0.002
Mpc$^{-1}$. The seven-year WMAP data \cite{Larson:2010gs} places
stronger bounds on $r$ than the five-year WMAP data \cite{WMAP5}.
In order to write down values that relate $n_s$ and $r$, we used
Eqs. (\ref{47}) and (\ref{54.1}), and the values $g_{*} = 100$,
$\gamma=1.5$, $\zeta_0 =(2/3)\zeta_{max}^{(1)}$, and  $m_{\phi}
=0.75\times10^{-5}$, $T = 2.5\times 10^{-6}$, $\Gamma_0=1.2\times
10^{-6}$, $\tau_0 = 3.73\times10^{-5}$, $\phi_0=0.3$ for $m=0,
n=0$; and $m_{\phi}=2.5\times10^{-5}$, $T = 1.75\times 10^{-6}$,
$\Gamma_0=3.58\times 10^{-6}$, $\tau_0 = 5.63\times10^{-5}$,
$\phi_0=0.6$ for $m=2, n=0$,   in Planck units \cite{HMB2004}.

Figure\ref{fig1} suggests that the pair of indices ($m = 2$, $n =
0$), corresponding to the right panel, is preferred over the other
pair of indices ($m = n = 0$), left panel. Likewise, it shows that
there is little difference between choosing $\lambda = 1$ or $
\lambda = 0.5$ as well as with the case of no viscosity, i.e.,
$\zeta_{0} = 0$.

Table \ref{tab:first-ex} indicates the value of the ratio $r$ for
$\lambda = 1$ and different choices of the pair of indices $m$ and
$n$ when the number of e-folds is $60$ and when it is $75$. Very
similar values (not shown) follow for $\lambda = 0.5$. All of them
can be checked with the help of Eqs. (\ref{47}) and (\ref{54.1}).

A comparison of the results shown in both Tables indicates that
only in the case of the pair ($m = n = 0$) with $N = 75$ (top and
bottom left panels in Fig. \ref{fig1}) viscosity makes a
non-negligible impact.

\begin{center}
\begin{table}
\begin{tabular}{|c|c|c|c|c|}
\hline
\hline
Panel in Fig. \ref{fig1} & $N$ & $ r $ & $N$ & $r$ \\
\hline
top left ($ m = n = 0$)& $60$ & $0.351$  & $75$ & $0.314$ \\
\hline
top right ($m = 2$, $n = 0$) & $60$ & 0.094 & $75$ & $0.074$ \\
\hline
\end{tabular}
\caption{\label{tab:first-ex} Results from first example with
$\lambda = 1$ (The results for $\lambda = 1/2$ are very similar).
Rows from top to bottom refers to panels of Fig. \ref{fig1} from
left to right.}
\end{table}
\end{center}

\begin{center}
\begin{table}
\begin{tabular}{|c|c|c|c|c|}
\hline
\hline
Panel in Fig. \ref{fig1} & $N$ & $ r $ & $N$ & $r$ \\
\hline
bottom left ($ m = n = 0$)& $60$ & $0.350$  & $75$ & $0.318$ \\
\hline
bottom right ($m = 2$, $n = 0$) & $60$ & 0.094& $75$ & $0.074$ \\
\hline \hline
\end{tabular}
\caption{\label{tab:first-ex} Results from first example with no
viscosity, i.e., $\zeta_{0}=0$.}
\end{table}
\end{center}

\subsection{Second example: effective Coleman-Weinberg potential}
As a second example we consider the effective Coleman Weinberg
potential near the critical point, for $T < T_{c}$ in the region
about $\phi_{B}$ \cite{berera}

\be V(\phi, T) = \alpha \, v_{1}(T) \, v_{2}(\phi)\, ,
\label{70}
\ee
\noindent where $v_{1} = k - T^{2}  \log (\delta / T) $ with
$\delta \sim m_{\rm Pl}$, and $v_{2} = (\phi - \phi_{B})^{2}$.  To
ensure that $V(\phi, T) \geq 0$ the condition $k \geq
\frac{\delta^{2}}{2 e}=k_{min}$ must be imposed.

The strength parameter adopts the form
\be Q = \frac{\Gamma(\phi,T)}{\sqrt{3\alpha \, v_{1} \,  v_{2}}}\,
, \label{71} \ee
whereas the slow-rolls parameters are given by
\be \epsilon = \eta = \frac{2}{(\phi-\phi_{B})^{2}}\, , \qquad
\beta = \frac{2 m }{\phi (\phi-\phi_{B})}\, , \qquad
b=\frac{T^{2}[1-2\log (\delta / T)]}{v_{1}}\, ,\qquad c = n\,
.\label{72} \ee
To enforce the smallness of the three first slow-rolls parameters,
the inflaton field must fulfill $\phi \gg \phi_{B} + \sqrt{2}$.

Thus, the energy density of the radiation fluid can be written as
\be
\rho_{\gamma}=\alpha \, (\phi-\phi_{B})^{2}\,  T^{2}\, (2\log
(\delta / T)-1)\, .
\label{73}
\ee

Following the procedure of the previous example, we find a lower
bound for $Q$, namely,
\be
Q > \frac{\alpha \mathcal{P}_{s}}{4}\left(\gamma
(\phi-\phi_{B})^{2}\frac{(2\log [\delta / T)-1]}{T^{2}} \, - \,
\frac{3}{\alpha} \frac{\zeta}{T^{3}}\right)\, ,
\label{74}
\ee
\noindent and using (\ref{73}), this relation simplifies to
\be
Q > \frac{\gamma \rho_{\gamma} \mathcal{P}_{s}}{4 T^{4}}
\left(1 - \frac{3}{\gamma} \, \zeta_{0} \, T \,
\rho_{\gamma}^{\lambda - 1} \right)\, .
\label{75}
\ee
\noindent Clearly, in this case as well,  viscosity increases the
range of $Q$.

\noindent The corresponding upper bound on $\zeta_{0}$, is
\be
\zeta_{0} < \zeta_{max}^{(2)} \equiv \frac{\gamma
\rho_{\gamma}^{1-\lambda}}{3 T}\, .
\label{76}
\ee
\begin{figure}[th]
\includegraphics[width=6.5in,angle=0,clip=true]{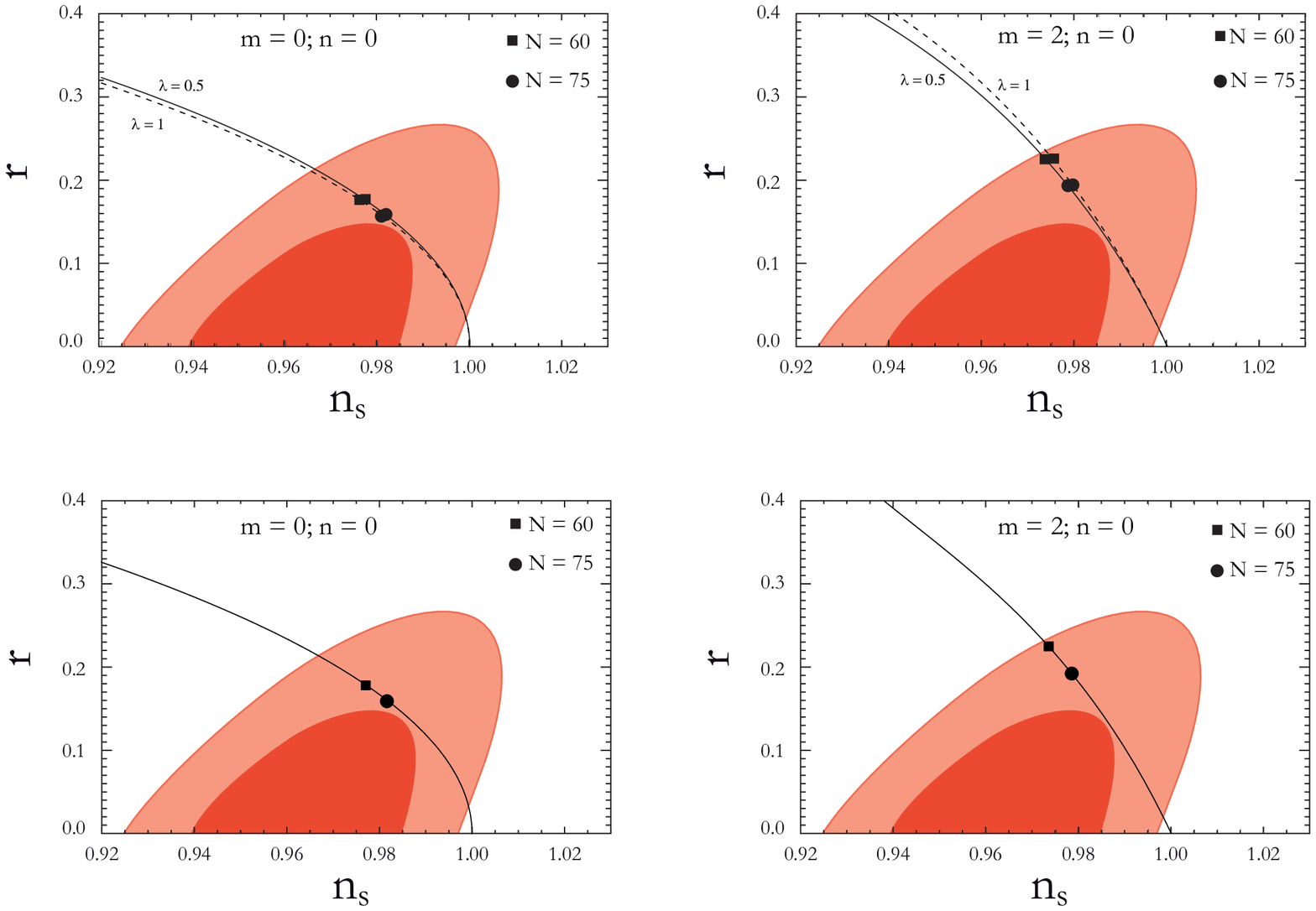}
\caption{Top row of panels: Plot of  the tensor-scalar ratio $r$
as a function of the spectral index $n_{s}$, for two values of the
$\lambda$ parameter in the case of example 2 (i.e., potential
(\ref{70})). Bottom row: Same as the top row but assuming no
viscosity ($\zeta_{0} =0$). In each panel  the 68$\%$ and 95$\%$
confidence levels set by seven-year WMAP experiment are shown. The
latter places severe limits on the tensor-scalar ratio
\cite{Larson:2010gs}.} \label{fig2}
\end{figure}
Figure \ref{fig2} illustrates the dependence of the tensor-scalar
ratio on the spectral index, from Eqs. (\ref{47}), (\ref{54.1}),
and (\ref{70}). We have used the following values: $\alpha = 3
\times 10^{-10}$, $\gamma = 1.5$, $\zeta_{0}
=(2/3)\zeta_{max}^{(2)}$,  $\phi_{B} = 2$, $\delta=1$,
$k=(3/2)\,k_{min}$  and  $\alpha =10^{-11}$, $T = 2.5\times
10^{-6}$, $\Gamma_0=1.2\times 10^{-6}$, $\tau_0 = 3.73 \times
10^{-5}$, $\phi_0=0.3$ for $m=0, n=0$; and $\alpha=10^{-8}$, $T =
1.75\times 10^{-6}$, $\Gamma_0=3.58\times 10^{-6}$, $\tau_0 = 5.63
\times 10^{-5}$, $\phi_0=0.6$ for $m=2, n=0$,
 in Planck units \cite{HMB2004}.

Figure \ref{fig2} hints that none of the two pair of indices is
clearly preferred over the other. On the other hand, it shows that
there is practically no difference between the choices $\lambda =
1$ and $\lambda = 1/2$.

\begin{center}
\begin{table}
\begin{tabular}{|c|c|c|c|c|}
\hline \hline
Panel in Fig. \ref{fig2} & $N$ & $ r $ & $N$ & $r$ \\
\hline
top left ($ m = n = 0$)& $60$ & $0.177$  & $75$ & $0.158$ \\
\hline
top right ($m = 2$, $n = 0$) & $60$ & $0.225$ & $75$ & $0.193$ \\
\hline \hline
\end{tabular}
\caption{\label{tab:second-ex} Results from the second example
with $\lambda = 1$ (very similar results, not shown, follow for
$\lambda = 1/2$). Rows from top to bottom refers to panels of Fig.
\ref{fig2} from left to right.}
\end{table}
\end{center}

\begin{center}
\begin{table}
\begin{tabular}{|c|c|c|c|c|}
\hline \hline
Panel in Fig. \ref{fig2} & $N$ & $ r $ & $N$ & $r$ \\
\hline
bottom left ($ m = n = 0$)& $60$ & $0.175$  & $75$ & $0.159$ \\
\hline
bottom right ($m = 2$, $n = 0$) & $60$ & $0.224$ & $75$ & $0.193$ \\
\hline \hline
\end{tabular}
\caption{\label{tab:second-ex2} Results from the second example in
the absence of viscosity, i.e., $\zeta_{0}=0$.}
\end{table}
\end{center}

Table \ref{tab:second-ex} indicates the value of the ratio $r$ for
$\lambda = 1$ and different choices of the pair of indices $m$ and
$n$ when the number of e-folds is $60$ and when it is $75$. Very
similar values (not shown) follow for $\lambda = 0.5$. All of them
can be checked with the help of Eqs. (\ref{47}) and (\ref{54.1}).

Again, a  the results of both Tables show, in this example, the
impact of viscosity is rather low.

\section{Conclusions}
We studied various aspects of warm inflationary universe models
when viscosity is taken into account. The latter (whether physical
or effective) is a very general feature in multiparticle and
entropy producing systems and, in the context of warm inflation,
it is of special significance when the rate of particle production
and/or interaction is high. In consequence we have focused on the
strong regime $Q \gg 1$ (in the weak regime viscosity is, at best,
negligible \cite{srd}).

Equation (\ref{27}) gives the necessary and sufficient condition
for the existence of
stable slow-roll solutions. Under this
condition, we got the same stability range obtained in the
no-viscous case, so long as $\sigma = -8/3$. In this sense, the
range of the slow-roll parameter $c$ decreases when $-8/3 < \sigma
< 0$, and increases when $\sigma < -8/3$.

We calculated a general expression for the spectral index, $n_{s}$
(Eq. (\ref{38})), that depends explicitly on viscosity through the
four  $p_{i}$ coefficients. The latter do not depend on the
slow-roll parameters $(\epsilon, \beta, \eta, {\rm and}\, b)$, as
shown by equations (\ref{39})-(\ref{42}).

In order to  further ensure the stability of the warm viscous
inflation, the slow-roll parameters must satisfy the following
conditions
\be
 \epsilon \ll 1+Q\, ,\quad |\beta| \ll 1 + Q\, , \quad |\eta|
\ll 1+Q \, ,
\nonumber
\ee
\noindent as well as the condition on the slow-roll parameter that
describes the temperature dependence of the potential, namely,
\be
|b| \ll \frac{(1-f)Q}{1+Q}\, .
\nonumber
\ee

To get explicit expressions we have considered two examples for
the scalar potential. In the first case, $V(\phi, T) = v_{1}
(\phi) \, + \, v_{2}(T)$ (Eq. (\ref{55})), it trivially follows
that $b$ vanishes; on the other hand, the stability condition
requires that the strength parameter fulfils
\be
Q > g_{*} \mathcal{P}_{s} \left(1 -
\frac{\zeta_{0}}{\zeta_{max}^{(1)}} \right)\, .
\nonumber
\ee
\noindent Hence, the viscous pressure increases the stability
range of $Q$ as $\zeta_{0} < \zeta_{max}^{(1)}$, where
$\zeta_{max}^{(1)}$ is given by Eq. (\ref{69}).

The second example is provided by the potential  $V(\phi, T) =
\alpha v_{1}(\phi) v_{2}(T)$, (Eq. (\ref{70})). In this case, the
stability condition for $Q$ is
\be Q > \frac{\gamma \rho_{\gamma} \mathcal{P}_{s}}{4
T^{4}}\left(1 - \frac{\zeta_{0}}{\zeta_{max}^{(2)}} \right)\, ,
\nonumber \ee
\noindent with $\zeta_{max}^{(2)}$ given by (\ref{76}). Again, the
range of stability of $Q$ augments in the presence of viscosity.

To bring in some explicit results we have taken the constraint
$n_{s}-r$ plane to first-order in the slow roll approximation. For
the first potential (Eq. (\ref{55})) we obtained that, when
$\lambda = 0.5$ and $\lambda = 1$, the model is consistent with
the WMAP seven year data for the pair of indices ($m= 2$, $n =
0$), see Fig. \ref{fig1}. Likewise, for the second potential
(Eq.(\ref{70})) all two choices of the ($m$, $n$) pair show
compatibility with the said data -see Fig.\ref{fig1}. The results
obtained indicate that the effect of viscosity on the value of $r$
is marginal. However, it cannot be discarded that future
experiments uncover it.

We should note that there are some properties in this model that
deserve further study. In particular, we did not address  the case
$n \neq 0$. In fact, when $\Gamma=\Gamma(\phi,T)$ a more detailed
and laborious calculation for the density perturbation would be
necessary in order to check the validity of expression (\ref{36}).
We intend to return to this point in the near future by following
an approach analogous to that of  Ref.\cite{Graham:2009bf}.

\acknowledgments{We are grateful to Mar Bastero-Gil for
conversations on the subject of this paper. This work was funded
by Comision Nacional de Ciencias y Tecnolog\'{\i}a through
FONDECYT Grants 1070306 (SdC) and 1090613 (RH and SdC), and by
DI-PUCV Grant 123787 (SdC) and 123703 (RH). Also partial support
came from the Spanish Ministry of Science and Innovation under
Grant FIS2009-13370-C02-01, and the ``Direcci\'{o} de Recerca de
la Generalitat" under Grant 2009SGR-00164.}


\begin{thebibliography}{9}
\bibitem{lyth1} A.R. Liddle and D.H Lyth, {\em Cosmological Inflation and Large-scale Structure}
(Cambridge University Press, Cambridge, 2000).
\bibitem{lyth2} D.H Lyth and A.R. Liddle, {\em The Primordial
Density Perturbation} (Cambridge University Press, Cambridge,
2009).
\bibitem{linde1} A. Linde, Phys. Lett. B \textbf{108}, 389 (1982).
\bibitem{linde2} A. Linde, {\em Particle Physics and Inflationary
Cosmology} (Harwood Academic, Chur, Switzerland, 1990).
\bibitem{berera-fang} A. Berera and L.Z. Fang, Phys. Rev. Lett.
\textbf{74}, 1912 (1995).
\bibitem{berera} A. Berera, Phys. Rev. Lett. \textbf{75}, 3218
(1995).
\bibitem{taylor2000} A. Taylor and A. Berera, Phys. Rev. D \textbf{62}, 083517 (2000).
\bibitem{salopek} D.S. Salopek and J.R. Bond, Phys. Rev. D
\textbf{42}, 3936 (1990).
\bibitem{parsons}A.R. Liddle, P. Parsons and J.D. Barrow, Phys. Rev.
D \textbf{50}, 7222 (1994).
\bibitem{henrique} H.P. de Oliveira and R.O. Ramos, Phys. Rev. D
\textbf{57}, 741 (1998).
\bibitem{moss2008} I.G. Moss and C. Xiong, JCAP 0811(2008)023.
\bibitem{weinberg1} S. Weinberg, Astrophys. J. \textbf{168}, 175
(1971).
\bibitem{weinberg2} S. Weinberg, {\em Gravitation and Cosmology} (J.
Wiley, New York, 1972).
\bibitem{zeldovich} Ya.B. Zeldovich, JETP Lett. \textbf{12}, 307
1970.
\bibitem{bei-lok} B.L. Hu, Phys. Lett. A \textbf{90}, 375
(1982).
\bibitem{barrow} J.D. Barrow, Nucl. Phys. B \textbf{310}, 743
(1988).
\bibitem{pla} W. Zimdahl and D. Pav\'{o}n, Phys. Lett. A \textbf{176}, 57 (1993).
\bibitem{josepedro} J.P. Mimoso, A. Nunes, and D. Pav\'{o}n, Phys.
Rev. D \textbf{73}, 023502 (2006).
\bibitem{srd} S. del Campo, R. Herrera, and D. Pav\'{o}n, Phys. Rev.
D \textbf{75}, 083518 (2007).
\bibitem{Graham:2009bf}
  C.~Graham and I.~G.~Moss,
  JCAP {\bf 0907}, 013 (2009).

\bibitem{Larson:2010gs} D.~Larson {\it et al.}, arXiv:1001.4635 [astro-ph.CO].
\bibitem{WMAP5} J.~Dunkley {\it et al.},
Astrophys.\ J.\ Suppl.\  {\bf 180}, 306 (2009);  G.~Hinshaw {\it
et al.}, Astrophys.\ J.\ Suppl.\  {\bf 180}, 225 (2009).
\bibitem{HMB2004} L. M. H. Hall, I. G. Moss, and A. Berera, Phys. Rev. D \textbf{69}, 083525 (2004).
\bibitem{Hoso} A. Hosoya and
M.-aki Sakagami, Phys. Rev. D 29, 2228 (1984).
\bibitem{zhang2009} Y. Zhang, JCAP 0903(2009)023.
\bibitem{1} W.~J.~Li, Y.~Ling, J.~P.~Wu and X.~M.~Kuang,
Phys. Lett.  B \textbf {687}, 1 (2010)
\bibitem{2} I. Brevik and O. Gorbunova, Gen. Rel. Grav. \textbf{37}, 2039 (2005); I.
Brevik, Int. J. Mod. Phys. D \textbf{15}, 767 (2006); X. M. Kuang
and Y. Ling, JCAP 10(2009)024.

\end{thebibliography}
\end{document}